# Nonlinear Electrodynamic Properties of Graphene and Other Two-Dimensional Materials

**Sergey MIKHAILOV**

Institute of Physics, University of Augsburg, Universitaetsstr. 1, Augsburg, D-86135, Germany
E-mail: sergey.mikhailov@physik.uni-augsburg.de

**Abstract:** Graphene, the first truly two-dimensional (one atom thin) material, possesses strongly nonlinear electrodynamic and optical properties. At low (microwave, terahertz) frequencies this results from the unique electronic property of graphene – the "relativistic", linear energy dispersion of its electrons and holes. At high (infrared, visible) frequencies its nonlinear response functions have resonances due to the inter-band transitions between the electron and hole bands. The position of these resonances on the frequency axis depends on the Fermi energy, or the charge carrier density, which can be varied by applying a gate voltage in the graphene field-effect transistor. This opens up the unique opportunities to electrically control the nonlinear graphene response, – the property which was not available in conventional (three-dimensional) nonlinear optical materials.

In this paper we give a short overview of the state of the art in this rapidly developing area of physics and present some of our recent results in this field. In particular, we discuss essential differences between the nonlinear response functions and parameters of traditional three-dimensional and new two-dimensional crystals.

# 1. Introduction

Graphene is a two-dimensional (2D) crystal consisting of a single layer of carbon atoms [1]. It was experimentally discovered in 2004 [2] and immediately attracted great interest of researchers. Graphene possesses a number of unique mechanical, thermal, electrical and optical properties. It is a semimetal, conducts electrical current and is transparent to visible light. Soon after its discovery it was reported about a whole family of 2D materials with all possible – dielectric, semiconducting and metallic – properties [3]. Built together, these 2D materials provide a basis for a completely new generation of electronic and optoelectronic devices with any desired electrical and optical properties. In fact, the discovery of graphene in 2004 opened, after the Stone Age, Bronze Age and Iron Age, a new era in the mankind history – the Age of 2D materials [4].

The carbon atoms in graphene occupy a 2D plane and are arranged in a honeycomb lattice. The spectrum of charge carriers in graphene can be calculated within the tight-binding approximation [5]. It consists of two bands, Fig. 1, which touch each other at the corners of the hexagonal Brillouin zone. Near these, so called Dirac points the spectrum of electrons and holes is linear,

$$E_{\pm}(\boldsymbol{k}) = \pm \hbar v_F |\boldsymbol{k}|, \quad (1)$$

where the electron wave-vector $\boldsymbol{k}$ is counted from the Dirac point, and the Fermi velocity $v_F \approx 10^8$ cm/s is a material parameter.

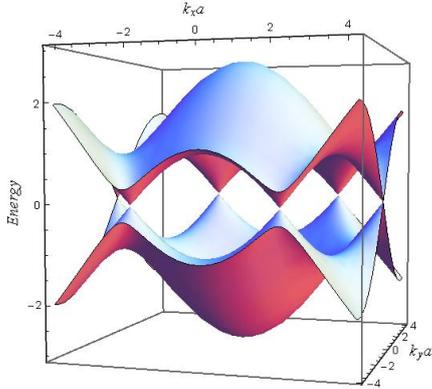

**Fig. 1.** The band structure of graphene calculated within the tight-binding approximation. The lower (valence) band is fully occupied in intrinsic graphene; the upper (conduction) band is empty. The bands touch each other at the corners of the hexagonal Brillouin zone. Near these points, which are referred to as the Dirac points, the spectrum of electrons and holes is linear.

It was theoretically predicted [6] that the linear energy dispersion (1) should lead to a strongly nonlinear electrodynamic response of this material. This can be seen from simple physical considerations. In contrast to massive particles, the velocity $\boldsymbol{v}_{\pm} = \partial E_{\pm}(\boldsymbol{k})/\partial \boldsymbol{p} = \pm v_F \boldsymbol{k}/|\boldsymbol{k}|$ of electrons with the spectrum (1) is not proportional to the momentum, therefore if an electric field $E_0 \cos(\omega t)$ acts on such electron, its momentum $\boldsymbol{p} = \hbar \boldsymbol{k}$ oscillates as $\sin(\omega t)$ but the velocity $\boldsymbol{v} \sim \text{sgn}[\sin(\omega t)]$ and hence the current contain higher frequency harmonics. All other nonlinear effects can evidently also be seen in graphene due to the same reason.

This prediction has been confirmed in numerous experiments performed in the last years, in which harmonics generation [7-9], four-wave mixing [10,11], Kerr effect [12-14] and other nonlinear phenomena have been observed in graphene at microwave through optical frequencies. In this paper which represents an extended version of our Conference presentation [15] we briefly overview the state of the art in the field and report about some of our recent results in the theory of nonlinear electrodynamics effects in graphene.

# 2. Linear Response of Graphene

Before starting to talk about the nonlinear graphene response we briefly outline its linear electrodynamic properties.

## 2.1. Dynamic Conductivity

The linear dynamic conductivity of graphene consists of two contributions [16-18],

$$\sigma^{(1)}(\omega) = \frac{e^2}{4\hbar}\left(S^{(1)}_{\text{intra}}(\Omega) + S^{(1)}_{\text{inter}}(\Omega)\right), \quad (2)$$

where the intra-band one,

$$S^{(1)}_{\text{intra}}(\Omega) = \frac{4}{\pi}\frac{i}{\Omega + i\Gamma_{\text{intra}}} \quad (3)$$

has the Drude form and the inter-band contribution

$$S^{(1)}_{\text{inter}}(\Omega) = \frac{i}{\pi}\ln\frac{2 - (\Omega + i\Gamma_{\text{inter}})}{2 + (\Omega + i\Gamma_{\text{inter}})} \quad (4)$$

has a step-like (logarithmic) feature in the real (imaginary) parts at $\Omega = 2$, Fig 2. Here $\Omega = \hbar\omega/|E_F|$, $\Gamma = \hbar\gamma/|E_F|$, $E_F$ is the Fermi energy, $\gamma$'s are the phenomenological relaxation rates (for the intra- and inter-band processes), and the formulas (3)-(4) are valid at zero temperature $T$=0. At a finite temperature the conductivity $\sigma^{(1)}(\omega, \mu, T)$ can be calculated according to the formula [19]

$$\sigma^{(1)}(\omega, \mu, T) = \frac{1}{4T}\int_{-\infty}^{\infty} \frac{\sigma^{(1)}(\omega, E_F, 0)}{\left[\cosh\left(\frac{\mu - E_F}{2T}\right)\right]^2} dE_F. \quad (5)$$

where $\mu$ is the chemical potential and $\mu_{T=0} = E_F$. The intra-band contribution does not essentially change when the temperature grows. As for the inter-band part, the finite $T$ effect leads to a broadening of the step-like and logarithmic features at $\Omega = 2$. Physically the features at $\hbar\omega = 2|E_F|$ are related to the inter-band absorption edge, Fig 3: the inter-band transitions at $\hbar\omega < 2|E_F|$ from empty states to empty states are forbidden; those at $\hbar\omega > 2|E_F|$ are allowed. This leads to the step-like absorption shown in Fig. 2.

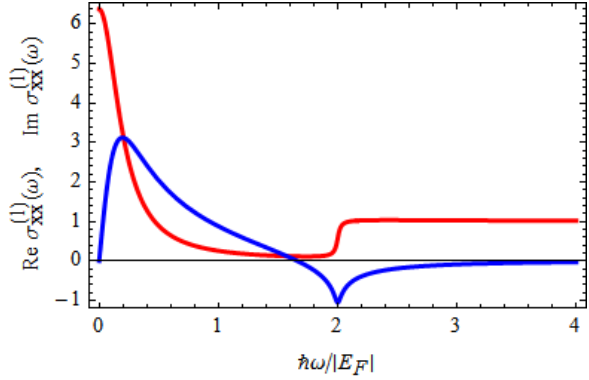

**Fig. 2.** The real (red) and imaginary (blue) parts of the conductivity (2) as a function of frequency at $T = 0$, $\Gamma_{intra} = 0.2$ and $\Gamma_{inter} = 0.02$.

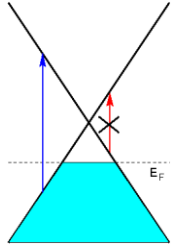

**Fig. 3.** The spectrum of electrons and holes near the Dirac point with Fermi energy $E_F < 0$. The vertical inter-band transitions with the frequency $\hbar\omega > 2|E_F|$ (blue arrow) are allowed; those with $\hbar\omega < 2|E_F|$ (red arrow) are forbidden.

## 2.2. Plasma Oscillations in Graphene

The spectrum of plasma waves in graphene was calculated in the Random Phase Approximation (RPA) in [20,21], Fig. 4. At low frequencies $\hbar\omega < 2|E_F|$ and small wave-vectors $q < k_F$ their spectrum is given by the square-root dispersion

$$\omega_p(\boldsymbol{q}) = \sqrt{\frac{2e^2 v_F}{\hbar\kappa}} \sqrt{\pi n_s q}, \qquad (6)$$

where $n_s$ is the electron (hole) density, and $\kappa$ the dielectric constant of the surrounding medium. At larger frequencies and wave-vectors the curve $\omega_p(\boldsymbol{q})$ enters the continuum of intra- and inter-band single-particle absorption (green and pink areas in Fig. 4), and the plasmons get damped. In addition to this, intra- and inter-band Landau damping, plasmons damp out due to impurity and phonons collisions – the effect not included in the RPA. In experiments, see, e.g., [22-24], the conditions $\hbar\omega < 2|E_F|$ and $q < k_F$ are usually satisfied, therefore Eq. (6) is sufficient for the description of 2D plasmons in graphene.

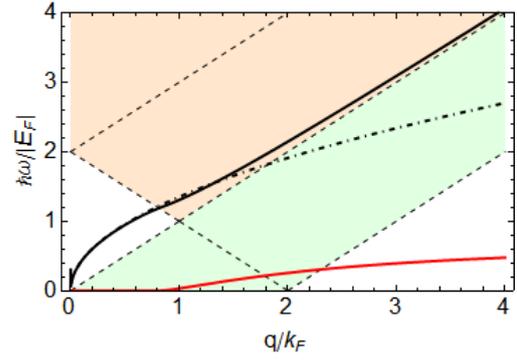

**Fig. 4.** The spectrum of plasma waves in graphene calculated in Refs. [20,21]. Black and red solid curves show the frequency $\omega'$ and the damping $\omega''$ of plasmons, the black dash-dotted curve is the low-frequency limit (6). In the colored areas the 2D plasmons have a finite damping.

## 3. Nonlinear Response of Graphene: Quantum Theory of the Third-Order Effects

The simple semiclassical picture of the nonlinear graphene response outlined in the Introduction is valid at low (microwave, terahertz) frequencies, $\hbar\omega < 2|E_F|$, when the inter-band transitions are not essential. A quantum theory valid at all frequencies was developed in [25-27]. In these papers the third-order conductivity tensor $\sigma^{(3)}_{\alpha\beta\gamma\delta}(\omega_1 + \omega_2 + \omega_3; \omega_1, \omega_2, \omega_3)$ was analytically calculated within the perturbative approach; here the first argument $\omega_1 + \omega_2 + \omega_3$ is the output signal frequency which equals the sum of three different input frequencies $\omega_1, \omega_2$ and $\omega_3$. Using the function $\sigma^{(3)}_{\alpha\beta\gamma\delta}(\omega_1 + \omega_2 + \omega_3; \omega_1, \omega_2, \omega_3)$ one can analyze a large number of different nonlinear effects.

### 3.1. Harmonics Generation

If a monochromatic radiation with the frequency $\omega$ is incident on a graphene layer the three input arguments of the third conductivity $\sigma^{(3)}_{\alpha\beta\gamma\delta}$ assume the values $\pm\omega$. The function $\sigma^{(3)}_{xxxx}(3\omega; \omega, \omega, \omega)$ determines the third harmonic generation. If the complex amplitude of the electric field vector of the incident wave is proportional to $\boldsymbol{E}_\omega \sim (\cos\vartheta, i\sin\vartheta)$ the third-harmonic current equals

$$\boldsymbol{j}_{3\omega} = \frac{1}{8}\sigma^{(3)}_{xxxx}(3\omega; \omega, \omega, \omega)\cos 2\vartheta \begin{pmatrix} \cos\vartheta \\ i\sin\vartheta \end{pmatrix}. \qquad (7)$$

Here $\vartheta$ is the ellipticity of the incident wave; $\vartheta = 0$ and $\pi/2$ correspond to the linearly, while $\vartheta = \pm\pi/4$ - to the circularly polarized waves. Equation (7) shows that the third harmonic intensity is maximal for the linearly polarized waves and vanishes for the circular polarization. The frequency dependence of the function $\sigma^{(3)}_{xxxx}(3\omega;\omega,\omega,\omega)$ which determines the third harmonic current is shown, at several temperatures, in Fig. 5. The curve corresponding to $T = 0$ shows resonances related to the inter-band absorption edge, Figure 3. In the third order there exist three such resonances, at $\hbar\omega = 2|E_F|$, $\hbar\omega = |E_F|$ and $\hbar\omega = 2|E_F|/3$, which correspond to the one-, two-, and three-photon absorption, respectively. These resonances can be tuned by the gate voltage which controls the Fermi energy and the electron or hole density in the sample. At room and higher temperatures the resonances get broader and are smeared out; here $T$ is the electron temperature which can essentially exceed the lattice temperature in the nonlinear optics experiments. In general the third harmonic generation efficiency falls down very quickly with frequency (as $\omega^{-3}$ at low and $\omega^{-4}$ at high frequencies).

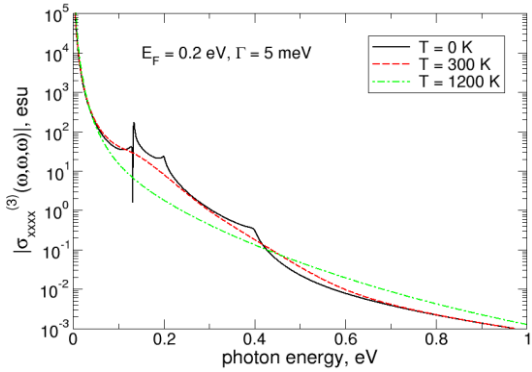

**Fig. 5.** The absolute value of the third conductivity $\sigma^{(3)}_{xxxx}(3\omega;\omega,\omega,\omega)$ of a single graphene layer as a function of the incident photon energy $\hbar\omega$ at the Fermi energy 0.2 eV and the effective relaxation rate 5 meV. Three resonant features at $T = 0$ K correspond to the one-, two-, and three-photon absorption at the absorption edge ($\hbar\omega = 2|E_F|$, $|E_F|$ and $2|E_F|/3$).

### 3.2. Kerr Effect and Saturable Absorption

The third-order response of graphene at the same frequency $\omega$ gives the Kerr and the saturable absorption effects. The corresponding contribution to the third-order current then has the form

$$\boldsymbol{j}_\omega = \frac{3}{8}\left[\sigma^{(3)}_{xxxx}(\omega;\omega,\omega,-\omega)\begin{pmatrix}\cos\vartheta\\ i\sin\vartheta\end{pmatrix} - \right. \quad (8)$$
$$-\sigma_{xyyx}3\omega;\omega,\omega,-\omega\sin 2\vartheta\sin\vartheta i\cos\vartheta.$$

Now one sees that the Kerr response is determined, at arbitrary $\vartheta$, by two components of the tensor $\sigma^{(3)}_{\alpha\beta\gamma\delta}$. The contribution from $\sigma^{(3)}_{xyyx}$ vanishes if the incident wave is linearly polarized but for elliptically and, in particular, for circularly polarized waves, this component contributes to the Kerr response. Measuring this response at the frequency $\omega$ one can thus determine not only $\sigma^{(3)}_{xxxx}(\omega;\omega,\omega,-\omega)$ but also $\sigma^{(3)}_{xyyx}(\omega;\omega,\omega,-\omega)$, see Section 6.2.

In general, there are four nonzero components of the tensor $\sigma^{(3)}_{\alpha\beta\gamma\delta}(\omega;\omega,\omega,-\omega)$. The two remaining components are related with $\sigma^{(3)}_{xxxx}(\omega;\omega,\omega,-\omega)$ and $\sigma^{(3)}_{xyyx}(\omega;\omega,\omega,-\omega)$ by the formulas [28]:

$$2\sigma^{(3)}_{xxyy}(\omega;\omega,\omega,-\omega) = \sigma^{(3)}_{xxxx}(\omega;\omega,\omega,-\omega) - \sigma^{(3)}_{xyyx}(\omega;\omega,\omega,-\omega), \quad (9)$$

$$\sigma^{(3)}_{xyxy}(\omega;\omega,\omega,-\omega) = \sigma^{(3)}_{xxyy}(\omega;\omega,\omega,-\omega). \quad (10)$$

Figure 6 shows the real and imaginary parts of $\sigma^{(3)}_{xxxx}(\omega;\omega,\omega,-\omega)$ at room temperature 300 K and at 1200 K. The real part of $\sigma^{(3)}_{xxxx}(\omega;\omega,\omega,-\omega)$ determines the nonlinear absorption in graphene. It is negative at all frequencies, which corresponds to a reduction of the linear absorption in strong electric fields, and has a broad resonant feature (at 300 K) around 0.4 eV (this corresponds to the condition $\hbar\omega = 2|E_F|$). The imaginary part of $\sigma^{(3)}_{xxxx}(\omega;\omega,\omega,-\omega)$ is positive at infrared and optical frequencies, has a broad resonance at $\hbar\omega = 2|E_F|$ and changes its sign at low frequencies. The positive sign of Im $\sigma^{(3)}_{xxxx}(\omega;\omega,\omega,-\omega)$ implies an *effective* negative refractive index of graphene corresponding to the self-defocusing Kerr nonlinearity [13-14], see Section 6.2 for further discussion of the experimental issues of the Kerr effect measurements.

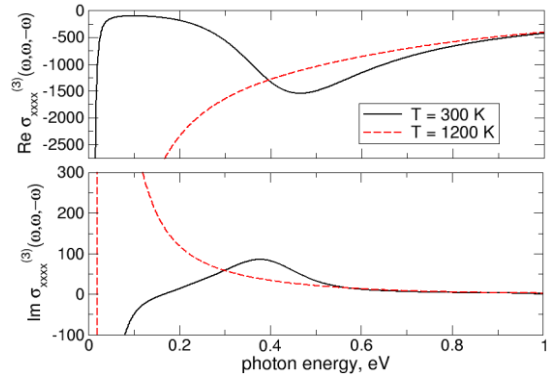

**Fig. 6.** The real and imaginary parts of the third conductivity $\sigma^{(3)}_{xxxx}(\omega;\omega,\omega,-\omega)$ as a function of $\hbar\omega$ at $E_F$=0.2 eV and $\Gamma$ = 5 meV.

The real and imaginary parts of $\sigma^{(3)}_{xyyx}(\omega;\omega,\omega,-\omega)$ are shown in Figure 7. They have similar frequency dependencies but smaller absolute values as compared to $\sigma^{(3)}_{xxxx}(\omega;\omega,\omega,-\omega)$.

### 3.3. Second Harmonic Generation from Graphene Driven by a Direct Current

Graphene is a centro-symmetric material; therefore under the action of a uniform external electromagnetic field it may produce only odd frequency harmonics. The central symmetry can be broken if a strong direct current is passed through the layer [29]. Figure 8 shows the function $\sigma^{(3)}_{xxxx}(2\omega;\omega,\omega,0)$ which determines the second harmonic generation from a single graphene layer irradiated by an electromagnetic wave with the frequency $\omega$ and driven by a strong direct current. Two resonances at $\hbar\omega = 0.2$ and $0.4$ eV correspond to the inter-band transitions at $\hbar\omega = |E_F|$ and $2|E_F|$. It is also noticeable that the absolute value of the function $\sigma^{(3)}_{xxxx}(2\omega;\omega,\omega,0)$ responsible for the second harmonic generation is several orders of magnitude larger than the third harmonic generation function $\sigma^{(3)}_{xxxx}(3\omega;\omega,\omega,\omega)$.

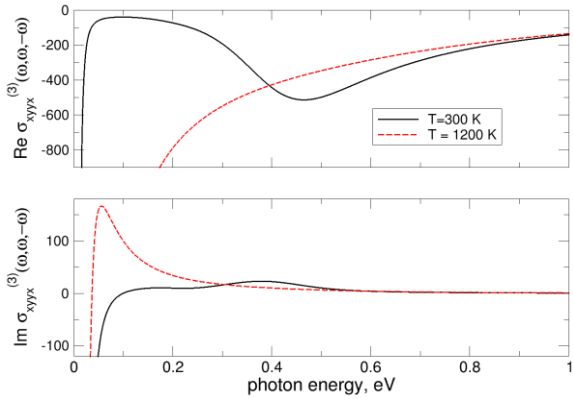

**Fig. 7.** The real and imaginary parts of the third conductivity $\sigma^{(3)}_{xyyx}(\omega;\omega,\omega,-\omega)$ as a function of $\hbar\omega$ at $E_F=0.2$ eV and $\Gamma = 5$ meV.

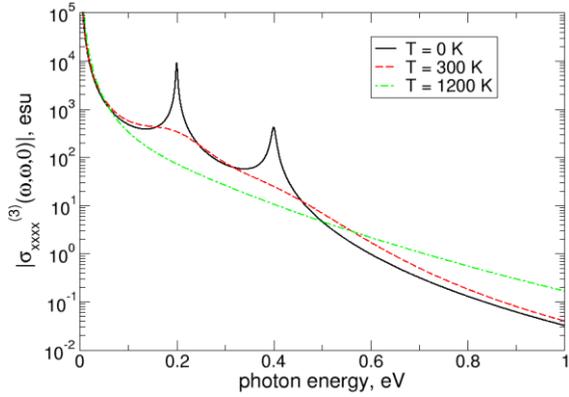

**Fig. 8.** The absolute value of the third conductivity $\sigma^{(3)}_{xxxx}(2\omega;\omega,\omega,0)$ of a single graphene layer as a function of $\hbar\omega$ at $E_F=0.2$ eV and $\Gamma = 5$ meV. The resonant features at $T = 0$ K correspond to $\hbar\omega = 2|E_F|$ and $|E_F|$.

Another opportunity to break the central symmetry of graphene was theoretically analyzed in [30,31]. In these papers the symmetry was broken by a finite wave-vector $q$ of the electromagnetic wave and different second-order nonlinear effects, such as the second harmonic generation, photon drag and difference frequency generation were considered.

### 3.4. Influence of Substrates on the Third Harmonic Generation in Graphene

The third harmonic generation efficiency from graphene essentially depends on the physical and geometrical properties of a substrate which the graphene layer lies on [32,33]. Different types of substrates were considered in these papers and their very strong influence on the harmonic generation efficiency was demonstrated, Figure 9. The results can be summarized as follows.

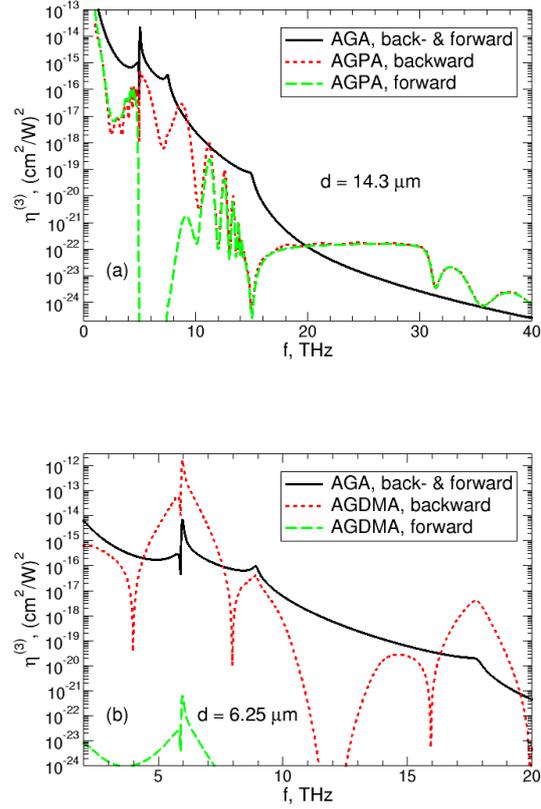

**Fig. 9.** Third harmonic generation efficiency $\eta^{(3)} = I_{3\omega}/I_\omega^3$ as a function of frequency: (a) in a structure air-graphene-polar dielectric-air (AGPA) and (b) air-graphene-dielectric-metal-air (AGDMA). Black curves show $\eta^{(3)}$ for isolated graphene (structure AGA) at $T = 0$ K; resonances are seen at the frequencies $\hbar\omega = 2|E_F|$, $|E_F|$ and $2|E_F|/3$. The transverse (TO) and longitudinal (LO) optical phonon frequencies in (a) lie at 15 and 30 THz respectively, $d$ is the dielectric thickness. The density of electrons is (a) $0.7\times 10^{11}$ cm$^{-2}$ and (b) $10^{11}$ cm$^{-2}$.

If graphene lies on a dielectric layer of thickness $d$ with a frequency independent refractive index $n$ the third harmonic generation efficiency $\eta^{(3)} = I_{3\omega}/I_\omega^3$ can be only smaller or equal to the efficiency $\eta^{(3)}$ in isolated graphene. The substrate does not influence $\eta^{(3)}$ only if the substrate thickness $d$ satisfies the interference condition $2d/\lambda = $ integer, where $\lambda$ is the wavelength of radiation in the dielectric. Otherwise the harmonic generation efficiency can be suppressed by *many orders of magnitude*, as compared to the isolated graphene layer.

If the substrate is made out of a polar dielectric with one or a few optical phonons resonances in the dielectric function $\varepsilon(\omega)$, the efficiency $\eta^{(3)}$ substantially increases in certain frequency intervals, Fig. 9(a), due to resonances at the TO and LO phonon frequencies. In addition, in the frequency range between $\omega_{TO}$ and $\omega_{LO}$ a flattening of the frequency dependence of $\eta^{(3)}$ is the case, Fig. 9(a). This effect can be useful for some applications since in the isolated graphene the efficiency $\eta^{(3)}$ very quickly falls down with the frequency, Section 3.1. The use of the polar dielectric substrate may help to achieve almost frequency independent third-order response of graphene. This effect is the case in the range of TO/LO phonon frequencies, i.e. at a few to a few tens of THz.

If the dielectric substrate is covered by a thin metallic layer on the back side, the efficiency of the third harmonic generation can be increased, in certain frequency intervals, by *more than two orders of magnitude* as compared to the isolated graphene, Figure 9(b). This happens if $2d/\lambda$ =integer+1/2 and is due to the interference of both the fundamental and the third harmonics in the dielectric substrate.

The correct choice of the substrate material and thickness is thus very important for successful device operation. The substrates may both suppress and increase the third harmonic generation efficiency $\eta^{(3)}$ by several orders of magnitude.

## 4. Nonperturbative Quasiclassical Theory of the Nonlinear Effects in Graphene

In Section 3 we discussed the results of the quantum theory of the third-order response of graphene which are valid at all frequencies ($\hbar\omega$ is smaller and larger than $2|E_F|$) but restricted by the third order Taylor expansion of the current in powers of the electric field. At low (microwave, terahertz) frequencies, $\hbar\omega < |E_F|$, one can develop a *nonperturbative* quasi-classical theory of the graphene response. At $\hbar\omega < |E_F|$ the interband transitions from the valence to the conduction band can be neglected and the nonlinear graphene response can be described by quasi-classical Boltzmann equation. If the scattering integral is written in the $\tau$-approximation form ($\tau$ is the momentum relaxation time) and the external ac electric field is uniform, Boltzmann equation can be solved exactly [34].

The theory [34] allows calculating the generation efficiency of all (odd) frequency harmonics in arbitrarily strong electric fields. Figure 10 shows as an example the field dependence of the generalized complex conductivity of the graphene layer $\sigma_{n\omega,\omega}(\omega\tau, F_\tau) \equiv \sigma_0 S_n(\omega\tau, F_\tau)$, which is defined as the ratio of the $n$-th harmonic of the current $j_{n\omega}$ to the 1st harmonic of the field $E_\omega$; here $\sigma_0$ is the static Drude conductivity and in Fig. 10 $n$=7. A noticeable result here is that in the perturbation-theory limit $F_\tau \ll 1$ the $n$–th harmonics current behaves as $j_{n\omega} \sim F_\tau^n$, i.e., falls down very quickly with $n$. In the strongly non-perturbative regime $F_\tau \gg 1$, the higher harmonics current decreases slowly with $n$, as $j_{n\omega} \sim 1/n$.

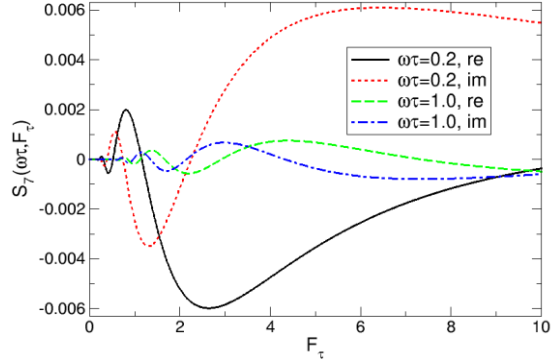

**Fig. 10.** The function $S_7(\omega\tau, F_\tau)$ proportional to the generalized conductivity $\sigma_{7\omega,\omega}(\omega\tau, F_\tau)$ as a function of the field strength $F_\tau = eE_\omega\tau/p_F$; here $p_F$ is the Fermi momentum.

The transmission $T$, reflection $R$ and absorption $A$ coefficients of the electromagnetic wave (frequency $\omega$) passing through a single graphene layer, have been also calculated in [34] at arbitrary values of the incident wave power. The frequency and power dependencies of $T$, $R$ and $A$ also depend on the ratio $\gamma_{rad}/\gamma$ of the radiative decay rate $\gamma_{rad}$ to the dissipative damping rate $\gamma=1/\tau$ [34]. At all values of the two parameters $\omega\tau$ and $\gamma_{rad}/\gamma$, the reflection decreases, while the transmission coefficient increases with the wave power. The absorption coefficient $A$ typically decreases (the saturable absorption effect) but in some cases (e.g. at $\gamma_{rad}/\gamma = 2$) the absorption $A$ first grows and then decreases with power [34].

The results [34] have been derived at $T = 0$. Their generalization to the finite temperature case can be done straightforwardly.

The $\tau$-approximation which was used in [34] is usually considered as insufficient for the description of the electromagnetic response at strong powers of radiation [35-37]. Under the strong irradiation the electron heating effects should be taken into account. If to generalize the theory [34] to the case of finite temperatures, it could then be used for analysis of experimental data by treating $T$ (the hot electron temperature) as a phenomenological fitting parameter. Such a work is still to be done.

## 5. Plasma Waves Nonlinearities in Graphene

If graphene is irradiated by an electromagnetic wave with a finite wave-vector $\mathbf{q}$ (in the direction parallel to the 2D layer plane), several new nonlinear effects are expected. Two of them are considered here.

## 5.1. 2D Plasmon Assisted Resonant Enhancement of the Second Harmonic

If $q$ is finite, the central symmetry of graphene is broken, see Section 3.3, and one can observe the second-order nonlinear effects, such the second (or, in general, even) harmonic generation, sum- and difference generation [30,31]. In addition, if the external field has spatial Fourier harmonics with the wave-vector $q$, one can excite plasma waves in the graphene layer, Section 2.2, if the frequency of radiation coincides with the 2D plasmon frequency (6). Near the resonance (6) the electric field acting on 2D electrons resonantly increases, which facilitates the observation of the nonlinear phenomena and leads, for example, to a resonant enhancement of the second harmonic intensity. Such effect was predicted in [38]. Figure 11 exhibits the second harmonic intensity as a function of the fundamental harmonic frequency [38]. Near the point $\omega \approx \omega_p(q)$ one sees a strong resonance due to the excitation of the 2D plasmons. At $\omega \approx 0.7\omega_p(q)$ one observes another, weaker resonance due to the resonant enhancement of the emitted second harmonic at the plasmon frequency, when $\omega^2 \approx \omega_p^2(q)/2$. The height of the resonances strongly depends on the plasmon quality factor $\omega_p \tau$. Thus, the 2D plasmon excitation helps to substantially increase the efficiency of the higher harmonics generation.

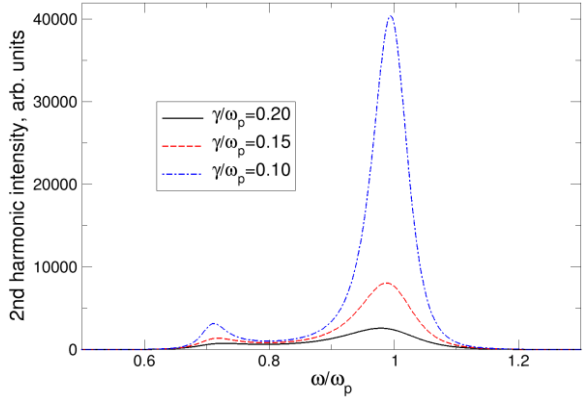

**Fig. 11.** The second harmonic intensity in graphene as a function of the radiation frequency measured in units of the plasma frequency $\omega_p(q)$. The parameter of the curves is $\gamma/\omega_p(q)$, where $\gamma = 1/\tau$ is the momentum relaxation rate.

It should be emphasized that, in order to observe the resonant enhancement of the second harmonic, shown in Fig. 11, one should really break the central symmetry of graphene. For example, the plasma waves are often observed in 2D electron systems by placing a grating in the vicinity of the 2D gas and measuring the far-infrared (FIR) transmission spectrum of such a structure [39]. The second harmonic will be excited in such a system only if the grating is intentionally made asymmetric.

## 5.2. Influence of Nonlinearities on the Plasma Waves in Graphene

The nonlinearity of the graphene response can modify the spectrum of 2D plasmons themselves. In Section 4 the generalized conductivity of graphene $\sigma_{n\omega,\omega}(\omega\tau, F_\tau)$ was discussed. Using the component $\sigma_{\omega,\omega}(\omega\tau, F_\tau)$ ($n = 1$) calculated in [34] one can find how the frequency and the linewidth of the 2D plasmon FIR transmission resonance are modified under the action of very strong external radiation.

In order to find the nonlinearity-induced modifications of the 2D plasmon spectrum we notice that the dispersion relation (6) follows from the formula

$$1 + \frac{2\pi i\,\sigma^{(1)}(\omega)}{\omega\kappa}q = 0, \qquad (9)$$

where $\sigma^{(1)}(\omega)$ is the linear conductivity of graphene. Replacing it by $\sigma_{\omega,\omega}(\omega\tau, F_\tau)$ we get the spectrum of 2D plasmons in graphene in the nonlinear regime [40]. Figure 12 illustrates the influence of the electric field strength on the excitation spectrum of graphene plasmons. A narrow weak-field plasma resonance centered at the plasma frequency $\omega_p$ shifts to lower frequencies and gets broader in the strong field regime. These predictions have been experimentally confirmed [41].

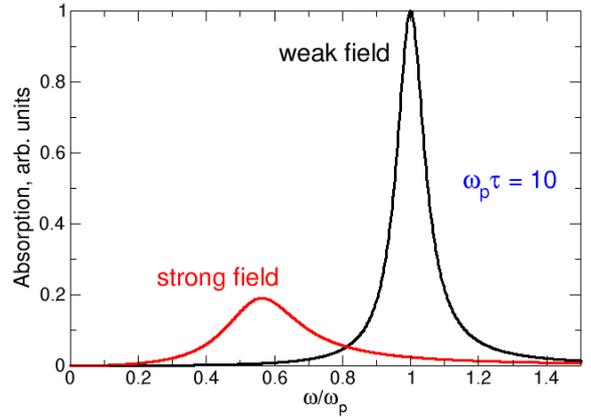

**Fig. 12.** Influence of the electric field strength on the shape of plasma resonances in a FIR absorption experiment in an array of narrow graphene stripes. The weak and strong field regimes correspond to $F_\tau \ll 1$ and $F_\tau = 15$.

Apart from the FIR transmission technique [22] the 2D plasmons in graphene can be studied by the scanning near-field optical microscopy (SNOM) [23,24]. In the SNOM experiments the frequency of the 2D plasmons is fixed and the wave-vector (in general, complex-valued) should be determined from the dispersion equation (9). In this case the wavelength $\lambda_p$ and the propagation length $L_p$ are relevant quantities which should be determined by the theory. Figure 13 shows a typical dependence of the length $L_p$ on the 2D plasmon frequency and the strength of the exciting electric field [40].

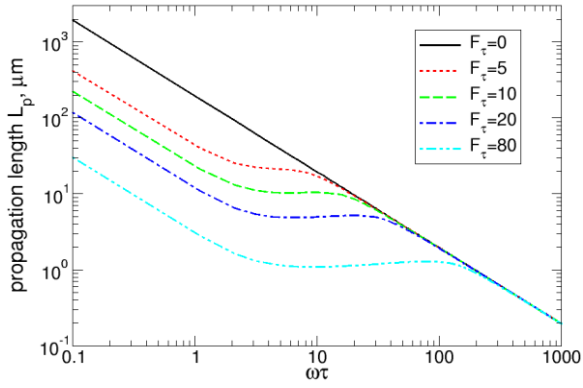

**Fig. 13.** The propagation length of 2D plasmons, as a function of frequency and the electric field strength, for typical graphene parameters.

## 6. Discussion of Recent Experiments and Some Special Features of the Nonlinear Electromagnetic Response of 2D Materials

### 6.1. Electrical Control of the Nonlinear Response of Graphene

In early experimental papers, e.g., in Refs. [7-14] and many other, the role of the parameter $\hbar\omega/2|E_F|$ was not fully understood. The experiments were performed on undoped graphene and at fixed incident wave frequencies. Theoretical predictions of the inter-band resonances [25-27] in the nonlinear graphene response functions opened the way to explore the unique property of graphene to *electrically control* nonlinear graphene properties. Very recently a number of experiments have been done aiming to demonstrate the ability to control the nonlinear optical response by the gate voltage in graphene based devices. The first demonstration of the electric tunability of the nonlinear graphene response was published in the four-wave mixing experiment [42]. The gate tunable third harmonic generation and the sum-frequency four-wave mixing were studied in [43]. Some other gate tunable results were shown in [44]. Results of these experiments perfectly agree with the predictions [25-27].

### 6.2. Theoretical Analysis of the Optical Kerr Effect Experiments

One of the nonlinear phenomena potentially important for applications is the optical Kerr effect (OKE), Section 3.2. It was measured in several publications, e.g. [12-14], by different methods, in particular, by the Z-scan and the optical heterodyne detection (OHD) techniques. The reported values of the *effective* nonlinear refractive index $n_2$ differed by orders of magnitude in different experiments, and even the sign of this quantity was found to be opposite in [12] and [13]; for further discussions and more references see [13]. In [13], where the Kerr effect was measured at the telecommunication wavelength ~1.6 μm by the OHD method the *effective* nonlinear refractive index $n_2$ was found to be about $-10^{-9}$ cm$^2$/W.

The OHD is a pump-probe technique in which both waves are incident on the graphene layer under different angles. The weaker wave (probe) has the polarization plane rotated by 45 degrees with respect to that of the pump wave, and can have a slight elliptic polarization. In [13] the intensities of both waves were modulated with different frequencies, $f_{pump}$ and $f_{probe}$, and the detector registered the transmitted through the graphene layer light at the sum of the modulation frequencies $f_{pump} + f_{probe}$.

In our recent paper [45] we have performed a detailed theoretical analysis of the OHD technique and, in particular, of the experiment [13]. We have shown that, in general, five nonlinear signals enter the detector, with the intensities proportional to $I_{pump}I_{probe}$, $I_{probe}^2$, $I_{pump}^2 I_{probe}$, $I_{pump} I_{probe}^2$, and $I_{probe}^3$. Each of these contributions can be measured at different modulation frequencies. The intensities of all these contributions are expressed via the real and imaginary parts of two independent components of the fourth-rank tensor $\sigma^{(3)}_{\alpha\beta\gamma\delta}(\omega;\omega,\omega,-\omega)$; for such independent components we have chosen in [45] $\sigma^{(3)}_{xxxx}(\omega;\omega,\omega,-\omega)$ and $\sigma^{(3)}_{xxyy}(\omega;\omega,\omega,-\omega)$. All other non-vanishing components of the $\sigma^{(3)}_{\alpha\beta\gamma\delta}$ tensor can then be found with the help of relations (9)-(10).

Thus, measuring the intensities of all five signals registered by the detector in the OHD method one can extract all information about the real and imaginary parts of the tensor $\sigma^{(3)}_{\alpha\beta\gamma\delta}(\omega;\omega,\omega,-\omega)$ which characterizes the Kerr effect in graphene. Together with the theoretical results [27] the paper [45] provides complete information about the theoretically expected dependencies of the OKE response on the radiation frequency and polarization, charge carrier density, temperature and scattering parameters of the studied 2D material.

Comparing our theoretical results with the experimental data of Refs. [12-14] we have found that the theory [45] gives the same sign of the *effective* nonlinear refractive index of graphene, as it was measured in [13-14]. The absolute values of $n_2$ estimated in the theory [45] and measured in the experiments reasonably agree with each other.

The last important issue that should be discussed is how the Kerr effect parameters measured by different methods (e.g., Z-scan or OHD) correlate with each other. It turns out that it does not make sense to compare the values of *effective* $n_2$ obtained by different methods. The reason is that the Kerr effect in graphene is characterized by two independent complex functions (see Section 3.2), e.g., $\sigma^{(3)}_{xxxx}(\omega;\omega,\omega,-\omega)$ and $\sigma^{(3)}_{xxyy}(\omega;\omega,\omega,-\omega)$, while the *effective* nonlinear refractive index $n_2$ obtained from the experiments is a real scalar quantity. The value of $n_2$ measured in the OHD experiment [13] at the wavelength ~1.6 μm is proportional to [45]

$$\text{Im}[\sigma_{xxxx}^{(3)}(\omega;\omega,\omega,-\omega) - \sigma_{xxyy}^{(3)}(\omega;\omega,\omega,-\omega)].$$

The value of $n_2$ extracted from the Z-scan experiment made with the linearly polarized light is proportional to $\text{Im}[\sigma_{xxxx}^{(3)}(\omega;\omega,\omega,-\omega)]$. These two quantities are thus completely different.

Is it possible to obtain not only $\sigma_{xxxx}^{(3)}$ but all nonzero components of the tensor $\sigma_{\alpha\beta\gamma\delta}^{(3)}$ in a Z-scan experiment? Yes, it is. To do this, the Z-scan experiment should be done using the elliptically polarized light. Then the quantity measured in such an experiment will be proportional to the linear combination of the $\sigma_{\alpha\beta\gamma\delta}^{(3)}$ components given by Eq. (8). The Z-scan technique can thus also be used to get full information about all nonzero components of the tensor $\sigma_{\alpha\beta\gamma\delta}^{(3)}(\omega;\omega,\omega,-\omega)$ describing the Kerr effect in 2D materials.

## 6.3. Special Features of the Nonlinear Response of 2D Materials

In the above discussion in Section 6.2 we have called the quantity $n_2$ the *effective* nonlinear refractive index emphasizing the word "effective". This is because such quantities as the linear or nonlinear "refractive index" ($n_0$ or $n_2$), the linear or nonlinear electric susceptibility ($\chi^{(1)}$ or $\chi^{(3)}$), which are widely used in experimental papers on the nonlinear response of graphene, are *incorrect* quantities to characterize the 2D materials. Indeed, the refractive index characterizes, by definition, the change of the velocity of light when it propagates in a medium different from vacuum. But graphene and other 2D materials are only one or a few atomic layers thick. How the light can propagate inside the medium which has a thickness of only one atom? This makes no sense. More formally, the refractive index is defined as a square root of the dielectric function, $n(\omega) = \sqrt{\varepsilon(\omega)}$. The definition of the dielectric function in macroscopic electrodynamics [46] implies a procedure of averaging microscopic electric fields over "physically infinitesimal" volume elements, which means that all sample dimensions should substantially exceed the inter-atomic distance. In graphene and other 2D materials the dielectric function and the refractive index (both linear and nonlinear) cannot be mathematically rigorously defined. Therefore their widespread use in experimental papers is inappropriate. The same is valid, in principle, for the widely used quantity $\chi^{(3)}$.

The physical quantities which correctly characterize the graphene response are the linear or third-order *conductivities*, $\sigma_{\alpha\beta}^{(1)}$ and $\sigma_{\alpha\beta\gamma\delta}^{(3)}$, defined as the 2D current (dimensionality A/cm) divided over the field. These quantities have a clear physical meaning and should therefore be used in the electrodynamics of 2D materials. In principle, the electric susceptibility $\chi$ can also be used if to define it properly, i.e., as the ratio of the dipole moment of a unit *surface* element of the 2D crystal (not of a unit volume element) divided by the electric field. Then the quantities $\chi^{(1)}$ and $\chi^{(3)}$ should be supplied by an additional index "2D", $\chi_{2D}^{(1)}$ and $\chi_{2D}^{(3)}$, to distinguish them from the conventional, three-dimensional quantities $\chi^{(1)}$ and $\chi^{(3)}$. The two-dimensional quantities $\chi_{2D}^{(1)}$ and $\chi_{2D}^{(3)}$ are measured, of course, in different units.

Another misunderstanding often arises in experimental works when, for example, the *effective* $\chi^{(3)}$ measured in a four-wave mixing (or Kerr) experiment is compared with the *effective* $\chi^{(3)}$ extracted from the third harmonic generation experiment. The third-order graphene conductivities $\sigma_{xxxx}^{(3)}(3\omega;\omega,\omega,\omega)$ and $\sigma_{xxxx}^{(3)}(\omega;\omega,\omega,-\omega)$, which characterize the harmonic generation and the Kerr effect respectively, are *very different functions*, see Figure 14. At optical frequencies $\hbar\omega > 2E_F$ they differ by many orders of magnitude. In addition, as seen from Fig. 14, both these functions vary greatly depending on the frequency and the electron density. Therefore it does not make sense to compare a nonlinear parameter which was measured at, say, $\lambda = 1.5$ μm with the same quantity, but measured in another paper at $\lambda = 0.5$ μm. Or, to discuss experimentally measured values of nonlinear parameters not indicating the Fermi energy of the electron density in the investigated sample. In addition, it should be noticed that the Γ-dependencies (the relaxation rates) of different nonlinear graphene parameters can also be very different. For example, at optical frequencies $\sigma_{xxxx}^{(3)}(3\omega;\omega,\omega,\omega)$ is practically insensitive to Γ while $\sigma_{xxxx}^{(3)}(\omega;\omega,\omega,-\omega)$ strongly depends on it, $\sim 1/\Gamma^2$.

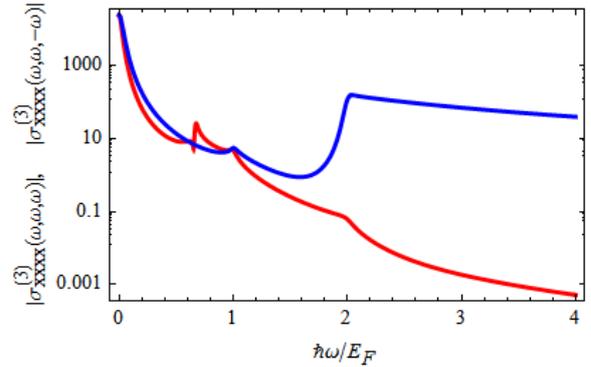

**Fig. 14.** The absolute values of the complex functions $\sigma_{xxxx}^{(3)}(3\omega;\omega,\omega,\omega)$ (red curve) and $\sigma_{xxxx}^{(3)}(\omega;\omega,\omega,-\omega)$ (blue curve) which characterize the harmonic generation and the Kerr effect respectively, as a function of $\hbar\omega/E_F$ at $T = 0$. Notice the logarithmic scale of the vertical axis.

All said in this Section is important for the appropriate interpretation of experimental results and the unambiguous understanding of nonlinear graphene physics.

## 7. Summary

We gave a brief overview of some selected theoretical and experimental results from the rapidly developing area of modern graphene physics – the nonlinear graphene electrodynamics and optics. Due to its unique physical properties, namely, the linear electron energy dispersion, graphene demonstrates the strongly nonlinear electromagnetic response at low (microwave, terahertz) frequencies. At higher, mid- and near-IR frequencies, it has strong resonances due to the inter-band electronic transitions. The unique feature of graphene is that the position of these resonances substantially depends on the electron density which can be tuned by the gate voltage in graphene based devices. All this opens up great opportunities for creation of electrically controlled nonlinear devices for microwave-, terahertz-, infrared- and optoelectronics, photonics and plasmonics.

## Acknowledgements

This work was funded by the European Union's Horizon 2020 research and innovation programme Graphene Core 2 under Grant Agreement No. 785219.